\documentclass[aps,pra,10pt,showpacs,twocolumn,superscriptaddress,longbibliography]{revtex4-1}
\usepackage{graphicx}
\usepackage[usenames]{color}
\usepackage{amssymb,amsmath}
\usepackage{physics}
\usepackage{siunitx}
\usepackage{xcolor}
\usepackage{setspace} 
\usepackage{float} 
\usepackage{tabularx}
\usepackage[linkcolor=blue,citecolor=blue,colorlinks=true,urlcolor=blue]{hyperref}

\newcommand{\fig}[1]{Fig.~\ref{#1}}
\newcommand{\be}[1]{\begin{equation}\label{#1}}
\newcommand{\ee}{\end{equation}}
\usepackage{amsmath,esint}
\usepackage{comment}
\usepackage{lipsum}
\usepackage[toc]{appendix}

\begin{document}

\title{Singularity in electron-core potential as a gateway to accurate multi-electron ionization spectra in strongly driven atoms}

\author{A. Emmanouilidou}
\affiliation{Department of Physics and Astronomy, University College London, Gower Street, London WC1E 6BT, United Kingdom}
\author{M. B. Peters}
\affiliation{Department of Physics and Astronomy, University College London, Gower Street, London WC1E 6BT, United Kingdom}
\author{G. P. Katsoulis}
\affiliation{Department of Physics and Astronomy, University College London, Gower Street, London WC1E 6BT, United Kingdom}

\begin{abstract}
We demonstrate a general three-dimensional semiclassical model  as a powerful technique for the study of correlated multi-electron escape in atoms driven by  infrared laser pulses at intensities where electron-electron correlation prevails. We do so in the context of triple ionization of strongly driven Ne. 
We show that a drawback of other current quantum mechanical and classical  models of triple ionization is that they soften the Coulomb potential of each electron with the core.
The model we employ  fully accounts for the singularity in the Coulomb potentials of  a recolliding electron with the core and a bound electron with the core as well as for 
the interaction of a recolliding with a bound electron. Our model treats approximately only the interaction between bound electrons through the use of 
effective potentials. These effective potentials ensure that no artificial autoionization takes place as a result of the full treatment of the electron-core potential.
We demonstrate the accuracy of our model by obtaining triple ionization distributions of the sum of the final electron momenta which we find to be in very good agreement with experiments.  Also, we explain the main features of these momenta distributions in terms 
of the prevalent pathways of correlated three-electron escape in Ne.  We also show that the different ionization pathways prevailing in  three-electron escape in  strongly driven Ne versus  Ar give rise  to different  momenta distributions in these two atoms.

 \end{abstract}
\date{\today}

\maketitle
 In atoms driven by intense and infrared  laser pulses, nonsequential multi-electron ionization (NSMI)  is a fundamental process  governed by electron-electron correlation \cite{knee1983}.  While nonsequential double ionization (NSDI) has been studied extensively both theoretically and experimentally \cite{Becker_review_2008,Ivanov_review_2009},  three-dimensional (3D) quantum mechanical studies  still remain quite challenging \cite{PhysRevLett.96.133001,Scrinzi_nsdi_2016,Scrinzi_nsdi_2020}. For nonsequential triple ionization (NSTI), only few  theoretical studies exist,  mostly formulated in the dipole approximation.   For NSTI, most studies employ lower dimensionality classical  \cite{PhysRevA.64.053401,PhysRevLett.97.083001} and quantum mechanical \cite{PhysRevA.98.031401,Efimov:21} models to reduce  the complexity  and  computational resources required. However, lower dimensionality results in a non accurate description of electron-electron interaction    during triple ionization. Currently,  only classical or semiclassical 3D models of NSTI are available  \cite{PhysRevLett.97.083001,Zhou:10,Tang:13,PhysRevA.104.023113,PhysRevA.105.053119}.  Here, we argue that the main disadvantage of  available  classical  and quantum models of NSTI is their softening of the interaction  of each electron with the core. This results in ionization spectra that differ from  experimental ones obtained, for instance,  for driven Ne and Ar \cite{PhysRevLett.84.447,Rudenko_2008,PhysRevA.86.043402,Herrwerth_2008,Zrost_2006,PhysRevLett.93.253001,SHIMADA2005221}.

Concerning NSTI,  for quantum mechanical models, softening the  Coulomb potential of each electron with the core  affords  a computationally tractable problem. For classical and semiclassical models, the reason is  fundamental and concerns unphysical autoionization. Classically there is no lower energy bound. Hence,  when a bound electron undergoes a close encounter with the core, the singularity in the Coulomb potential allows this electron to acquire a very negative energy. This can lead to the artificial escape of another bound electron through the Coulomb interaction between bound electrons.
 To avoid this,   most classical and semiclassical models of NSTI  soften the Coulomb potential  \cite{PhysRevLett.97.083001,Zhou:10,Tang:13}  or   add Heisenberg potentials  \cite{PhysRevA.21.834} (effective softening) to mimic the Heisenberg uncertainty principle and prevent  each electron from a close encounter with the core  \cite{PhysRevA.104.023113,PhysRevA.105.053119}.  
 
 However, softening the Coulomb potential fails to accurately describe electron scattering from the core \cite{Pandit2018,Pandit2017}. Indeed, the ratio of the scattering amplitude for the soft-core potential over the one
    for the Coulomb potential decreases exponentially with increasing momentum transfer \cite{Pandit2018,Pandit2017}. For recollisions \cite{Corkum_1994}, this implies that  soft potentials are quite inaccurate for high energy recolliding electrons that backscatter.
    Hence, it is no surprise that classical models that include the singularity in  the Coulomb electron-core potential
     result in accurate double ionization spectra.
     Indeed, with a classical model  for driven two-electron atoms \cite{PhysRevA.78.023411}, the predecessor of the model of NSMI discussed here, we have shown  that backscattering of the recolliding electron from the core  gives  rise to the finger-like structure in the two electron correlated   momenta of driven He \cite{PhysRevLett.96.133001,PhysRevLett.99.263002,PhysRevLett.99.263003}. We have  also obtained   double ionization spectra in very good agreement  with an \textit{ab initio} quantum mechanical calculation for driven   He \cite{Emmanouilidou_2011} and  with an experiment   for Ar driven by near-single cycle laser pulses \cite{ChenA2017Ndiw}.  We have  also identified the striking slingshot-NSDI  mechanism where the exact treatment of the  electron-core interaction is of paramount importance \cite{Slingshot}.
 
 Here, we provide a general 3D classical model of NSMI developed in the nondipole framework.   The main premise in our model is that
  two interactions are most important during a recollision and hence are treated exactly. Motivated by the above mentioned  studies \cite{Pandit2018,Pandit2017,PhysRevA.78.023411,ChenA2017Ndiw,Slingshot}, we account for the singularity in the Coulomb potential between  each electron,  bound or quasifree, and the core. Quasifree refers to a recolliding electron or an electron escaping to the continuum.  Also,  the  Coulomb potential between each pair of a  quasifree and a bound electron and hence the transfer of energy from a quasifree  to a bound electron
  is treated exactly.   Using this model, for nonsequential triple ionization of strongly driven Ne, we obtain   triple ionization spectra in excellent agreement with experiment \cite{Rudenko_2008}.
   
  Accounting for the singularity in electron-core interactions can lead  to  unphysical autoionization through energy transfer between bound electrons.  To avoid this, we use  effective Coulomb potentials to account for  the interaction of a bound-bound  electron pair---ECBB. That is, we approximate the  energy transfer  from a bound to a bound  electron. 
  Hence, we expect that the ECBB model will be more accurate  for laser pulse parameters where multi-electron ionization  due to transfer of energy between electrons in excited states after recollisions plays less of a role.
    A sophisticated aspect of the ECBB model involves deciding during time propagation whether an electron is quasifree or bound. That is, 
  we decide on the fly if the full or effective Coulomb potential 
  describes 
   the interaction between a pair of electrons. To do so,  we use a set of simple criteria detailed below.

We demonstrate the accuracy of the ECBB model in the context  of correlated three-electron escape in strongly driven Ne. We show that  the z-component of the sum of the final  electron momenta   has  excellent agreement with experiment \cite{Rudenko_2008}. Here, the electric field is linearly polarized along the $z$ axis.
 The ECBB model has been previously used to study  triple ionization of strongly driven Ar \cite{Agapi3electron}. However, the striking agreement with experiment for strongly driven Ne unveils   the ECBB model as  a  powerful  technique for studying
 correlated multi-electron ionization in  driven atoms. Moreover, we interpret the features of the z-component of the sum of the final electron momenta in terms of the main 
 recollision pathways for driven Ne and Ar. The differences in the  ionization spectra of the two atoms are found to be due to direct  pathways prevailing  triple ionization of Ne.


We employ the ECBB  model  \cite{Agapi3electron} to compute triple and double ionization observables of driven Ne. In what follows, TI refers to NSTI and DI to NSDI. One electron tunnel ionizes through the field-lowered Coulomb barrier at time $t_0$. In our previous studies  of double ionization \cite{PhysRevA.78.023411,Slingshot,Emmanouilidou1,Emmanouilidou2} tunneling occurs with a rate described by the quantum-mechanical Ammosov-Delone-Krainov formula  \cite{Landau,Delone:91}.  Here, using the same formula,  hence the term semiclassical model, we obtain a rate that also accounts for depletion of the initial ground state, see  \cite{sup_mat}.  We find $t_0,$ using importance sampling \cite{ROTA1986123} in the time interval [-2$\tau$, 2$\tau$] where the electric field is nonzero; $\tau$ is the full width at half maximum of the pulse duration in intensity. The exit point of the recolliding electron along the direction of the electric field is obtained  analytically using parabolic coordinates \cite{HUP1997533}. The electron momentum along the electric field is set equal to zero, while the transverse one is given by a Gaussian distribution \cite{Delone:91,Delone_1998,PhysRevLett.112.213001}. 
For the initially bound electrons, we employ   a microcanonical distribution  \cite{Agapi3electron}.

In the ECBB model, we fully account for the magnetic field of the laser pulse, i.e. the magnetic field component of the Lorentz force, as well as  the motion of the core and the three electrons. The four-body Hamiltonian  is 
\begin{equation}\label{Hamiltonian_effective}
\begin{aligned}
&H = \sum_{i=1}^{4}\frac{\left[\mathbf{\tilde{p}}_{i}- Q_i\mathbf{A}(y,t) \right]^2}{2m_i}+\sum_{i=2}^{4}\frac{Q_iQ_1}{|\mathbf{r}_1-\mathbf{r}_i|} \\
&+\sum_{i=2}^{3}\sum_{j=i+1}^{4} \left[ 1-c_{i,j}(t)\right]\frac{Q_iQ_j}{|\mathbf{r}_i-\mathbf{r}_j|} +\sum_{i=2}^{3}\sum_{j=i+1}^{4}c_{i,j}(t) \\\Big[ 
& V_{\mathrm{eff}}(\zeta_j(t),|\mathbf{r}_{1}-\mathbf{r}_{i}|) + V_{\mathrm{eff}}(\zeta_i(t),|\mathbf{r}_{1}-\mathbf{r}_{j}|)\Big],
\end{aligned}
\end{equation}
where $Q_i$ is the charge, $m_i$ is the mass, $\mathbf{r}_{i}$ is the position vector and $\mathbf{\tilde{p}}_{i}$ is the canonical momentum vector of particle $i$. The mechanical momentum $\mathbf{p}_{i}$ is given by
\begin{equation}
\mathbf{p}_{i}=\mathbf{\tilde{p}}_{i}-Q_i\mathbf{A}(y,t).
\end{equation}
The effective Coulomb potential that an electron $i$ experiences at a distance $|\mathbf{r}_{1}-\mathbf{r}_{i}|$ from the core (particle 1 with $Q_{1}=3$),  due to the charge distribution of electron $j$ is equal to 
\begin{equation}
V_{\mathrm{eff}}(\zeta_j,|\mathbf{r}_{1}-\mathbf{r}_{i}|) =  \dfrac{1-(1+\zeta_j| \mathbf{r}_{1}-\mathbf{r}_{i}|)e^{-2\zeta_j| \mathbf{r}_{1}-\mathbf{r}_{i}|}}{| \mathbf{r}_{1}-\mathbf{r}_{i}|}, 
\end{equation}
with $\zeta_j$ the effective charge of particle $j$ \cite{PhysRevA.40.6223,Agapi3electron}. When $ \mathbf{r}_{i}\rightarrow\mathbf{r}_{1}$, the effective potential   is equal to $\zeta_j$. This ensures a finite transfer of energy   between bound electrons $i$ and $j$ and hence  that  no artificial autoionization occurs.   The functions $c_{i,j}(t)$ determine at time $t$ during propagation whether  the full Coulomb or  effective $V_{\mathrm{eff}}(\zeta_i,|\mathbf{r}_{1}-\mathbf{r}_{j}|)$ and $V_{\mathrm{eff}}(\zeta_j,|\mathbf{r}_{1}-\mathbf{r}_{i}|)$ potentials describe the interaction between electrons $i$ and $j$  \citep{Agapi3electron}. The effective potentials are activated only when both electrons are bound. 
 
 We determine on the fly whether an electron is quasifree or bound using the following simple criteria. A quasifree electron can  transition to bound following a recollision. Specifically, 
 once the quasifree electron has its closest encounter with the core, this electron transitions to bound if its position  along the $z$ axis is  influenced more by the core than the electric field. Also, a bound electron  transitions to quasifree due to transfer of energy  during a recollision  or from the laser field.  In the former case, this transition occurs if the potential energy of  this bound electron with the core is constantly decreasing. In the latter case,  if  the energy of the bound electron  becomes  positive and remains positive it transitions to quasifree. The criteria are discussed in detail and  illustrated in \cite{sup_mat}.

Details of  how we accurately account for the Coulomb singularity and the leapfrog technique we employ to solve  Hamilton's equations of motion are given in Ref. \cite{Agapi3electron}.
We stop the  propagation when the energy of each particle converges. We   label the trajectory as  triply or  doubly ionized  if three or two electrons have  positive energy,  and compute the  TI and DI probabilities out of all events.

We use a vector potential of the form
\begin{equation}\label{eq:vector_potential}
\mathbf{A}(y,t) = -\frac{E_0}{\omega}\exp \left[ - 2\ln (2)\left( \frac{c t - y}{c \tau} \right)^2 \right]   \sin ( \omega t  - k y) \hat{\mathbf{z}},
\end{equation}
where $k=\omega/c$ is the  wave number of the laser field.  The direction of the vector potential and the electric field, $\mathbf{E}(y,t) = -\frac{\partial\mathbf{A}(y,t)}{\partial t}$, is along the $z$ axis, while the direction of light propagation  is along the $y$ axis.  The magnetic field, $ \mathbf{B}(y,t) = \grad \times \mathbf{A}(y,t)$, points  along the $x$ axis.  The pulse duration is $\tau = 25$ fs, while  the wavelength is  800 nm. For Ne, we consider intensities 1.0, 1.3 and $\mathrm{1.6 \; PW/cm^2 }$. For Ar, previously studied in Ref.  \cite{Agapi3electron}, we consider  only  $\mathrm{0.4 \; PW/cm^2 }$. The  highest intensities  considered here,  $\mathrm{1.6 \; PW/cm^2 }$ for  Ne and $\mathrm{0.4 \; PW/cm^2 }$ for  Ar, are chosen such that  the probability for a second electron to tunnel ionize solely due to the laser field is very small  \cite{sup_mat}. 
 Hence,  electron-electron correlation prevails in TI and DI, with  the bound electrons ionizing only  due to recollisions.
The smaller intensity for Ar is consistent with its smaller first ionization potential.


Here, we compare the results obtained with the ECBB model both with experiment and with the semiclassical 3D model that employs Heisenberg potentials (H model), see \cite{sup_mat}. This potential depends on a parameter $\alpha$, with a   large value  restricting more the phase space an electron can access around the core \cite{Agapi3electron,sup_mat}. Hence, the H model  results in an effective softening of  the electron-core potential. We do not compare with classical models that explicitly soften the Coulomb potential. The reason is that a previous  study of NSDI  in Ar   \cite{Sarkadi_2020} has shown   that the H model and the model that includes the Coulomb singularity  \cite{PhysRevA.78.023411,Emmanouilidou_2011,ChenA2017Ndiw}   better agree with experiment.

\begin{figure}[t]
\centering
\includegraphics[width=\linewidth]{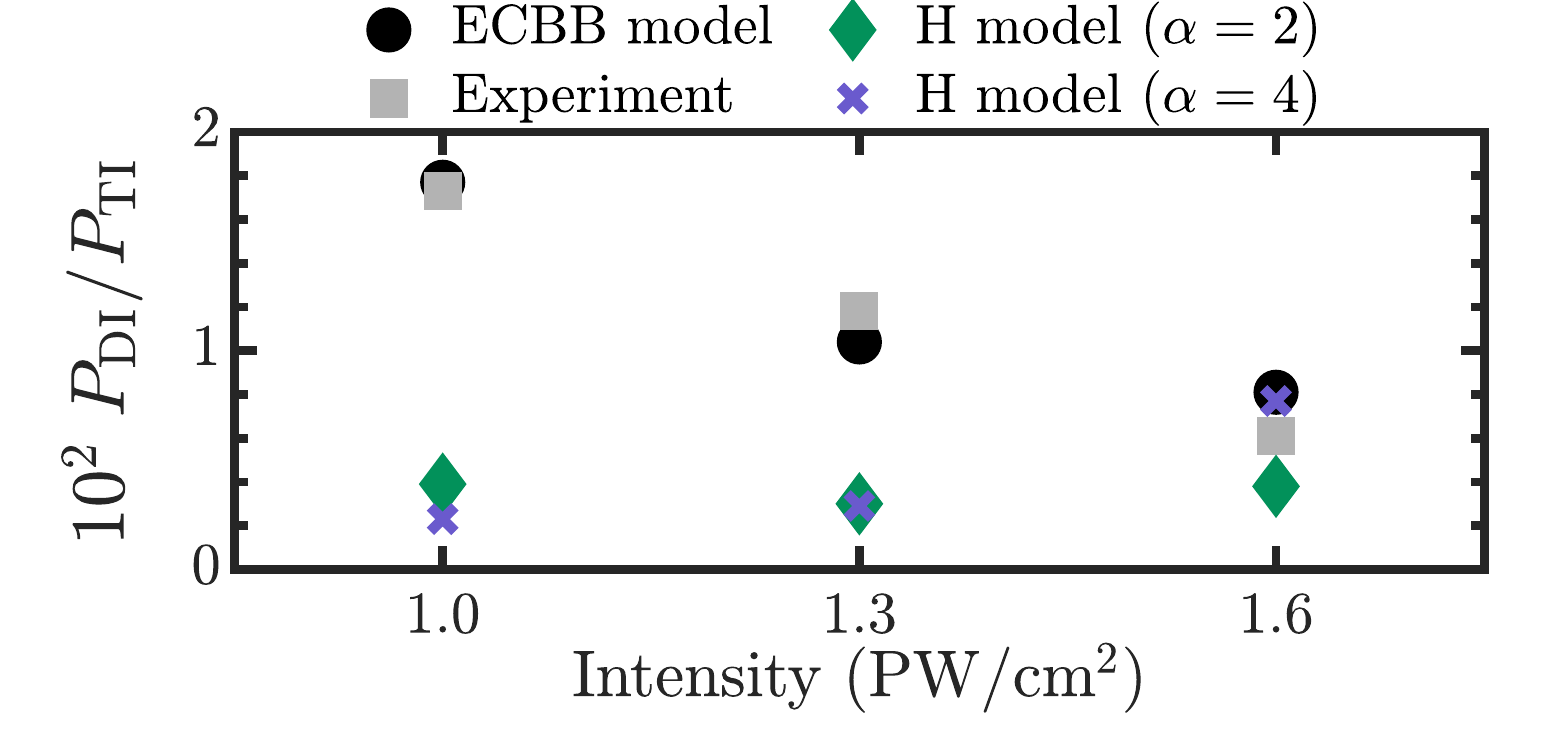}
\caption{For Ne, ratio of DI to TI  probability obtained with the ECBB model (black circles), the H model  and experiment (grey squares) \cite{Rudenko_2008}.}\label{fig:ratio_of_probabilities}
\end{figure}

 In  \fig{fig:ratio_of_probabilities}, for driven Ne, we compute the ratio of double to triple ionization probability and compare with  experiment \cite{Rudenko_2008} and  the H model.  For  all three intensities, 
 we find   the  probability ratio, $P_{\mathrm{DI}}/P_{\mathrm{TI}}$, obtained with the ECBB model (black circles) to be consistently close to experiment (grey squares). In contrast, the H model for  $\alpha=2$  and  $\alpha=4$  does not agree with experiment for 1.0 and $\mathrm{1.3 \; PW/cm^2 }$. Also, we find that the DI probability depends on the value of $\alpha$ for driven Ne \cite{sup_mat}, similar to our previous findings   for driven Ar  \cite{Agapi3electron}.   

\begin{figure}[t]
\centering
\includegraphics[width=\linewidth]{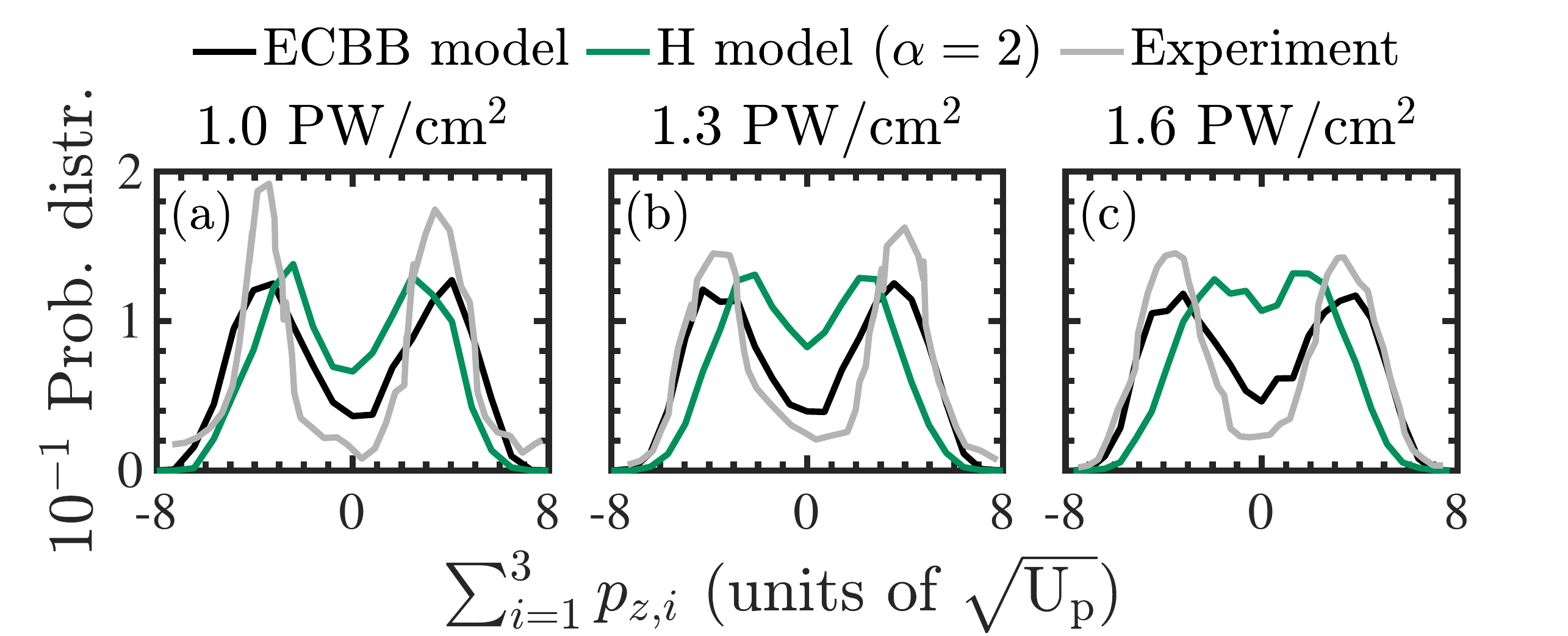}
\caption{For Ne, TI probability distributions of  the sum of $p_z$ obtained with the ECBB model (black lines), with   the H model (green lines) and measured experimentally  \cite{Rudenko_2008} (grey lines). Distributions are normalized to one. $U_{p}$ is the ponderomotive energy equal to $E_0^2/4\omega^2$.}\label{fig:SOM_TI}
\end{figure}

 Next, we  compute the TI probability distribution of the z-component  of the sum  of the final electron momenta, sum of $p_z$, see black lines
in \fig{fig:SOM_TI}. We compare with measurements (grey lines) \cite{Rudenko_2008}, smoothed in \fig{fig:SOM_TI},  and with the H model for $\alpha=2$ (green lines). We find the ECBB distributions to be doubly peaked at all intensities. With increasing intensity, the peaks become less pronounced with an increasing probability for the  sum of $p_z$  to be around zero. These features  agree well with experiment. Also, the ECBB distributions peak  at roughly the same values of the sum of the electron momenta as the experiment. This excellent agreement further illustrates the accuracy of the ECBB model.
 In contrast, the H model distributions have a significantly higher probability for the  sum of the final electron momenta to be around zero. Also, they are less wide compared to the ECBB model and experimental  distributions. 
 The difference   is  more pronounced at $\mathrm{1.6 \; PW/cm^2 }$, [\fig{fig:SOM_TI} (c)] with the H model distribution peaking around zero and the other two distributions  being doubly peaked.  This difference shows that in the H model  the effective softening of the interaction of the recolliding electron with the core  results in electrons escaping with lower energy. This  gives rise to  less wide distributions that have significant probability for the sum of $p_z$ to be around  zero.
 

 Next, using the ECBB model, we analyze the TI  events and identify the recollision pathways that prevail in the three-electron escape of driven Ne. In the Supplementary Material \cite{sup_mat}, we outline the algorithm we use to identify the recollision pathways. An electron is deemed 
 as ionizing soon after recollision if the difference between the recollision time and the ionization time is less than $t_{\text{diff}}=T/8$, where $T$ is the period of the laser pulse. During this time interval, the interpotential energy between the recolliding and a bound electron undergoes a sharp change.
  The recollision time is identified from the maxima in the interpotential energies between the recolliding and each of the bound electrons \cite{sup_mat}.
 The ionization time of  electron $i$  is defined as the time when  the compensated energy $\{[\mathbf{p}_{i}-\mathbf{A}(y,t)]^2+V(r_{i})\}$ of this electron becomes positive and remains positive thereafter \cite{Leopold_1979}. 

For  driven Ne, we find that  two  are the main recollision pathways  contributing to triple ionization, the direct   $(e^{\text{-}},3e^\text{-})$ and the delayed $(e^{\text{-}},2e^\text{-})$. For a recollision to take place, an electron tunnels out through the field-lowered Coulomb barrier \cite{Corkum_1994}. This  electron can then return to the parent ion to  recollide and transfer energy to the remaining electrons.
 In the direct pathway,   all three electrons ionize soon after recollision, i.e. there are three highly correlated electron pairs.  In the delayed $(e^{\text{-}},2e^\text{-})$ pathway, the recolliding electron transfers enough energy for only two electrons to ionize soon after recollision, while the other electron ionizes with a delay. Hence, there is only one highly correlated electron pair. 
  At all three intensities, we find that recollisions occur around a zero of the electric field and  a maximum of the vector potential, resulting in a large final electron momentum with magnitude $E_{0}/\omega=2\sqrt{U_{p}}$. In the direct pathway, all three electrons escape with large momenta $p_z$
  versus  two electrons  in the delayed $(e^{\text{-}},2e^\text{-})$ pathway, see also the correlated electron momenta  in \cite{sup_mat}.

\begin{figure}[t]
\centering
\includegraphics[width=\linewidth]{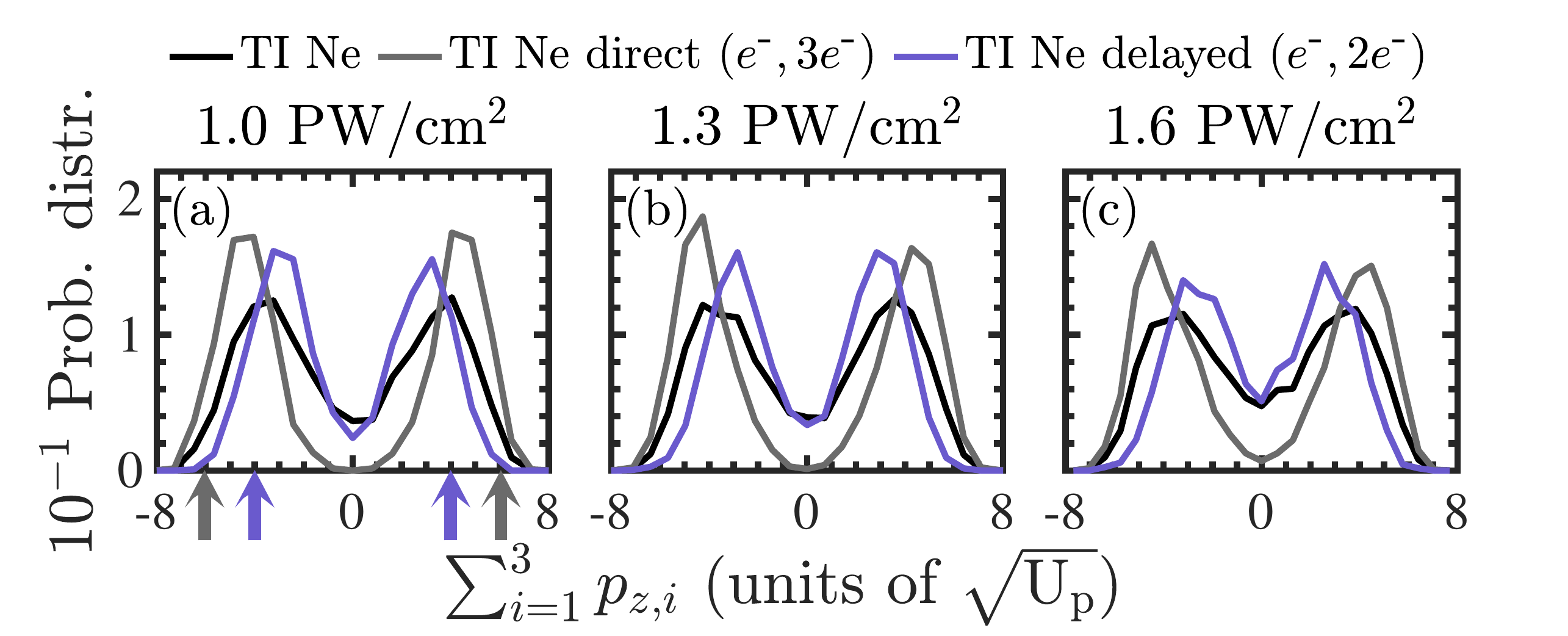}
\caption{For Ne, probability distributions of the sum of  $p_z$ for TI obtained with the ECBB model for all  (black), direct  (dark grey) and delayed  (blue) events. }\label{fig:SOM_TI_pathways}
\end{figure}
Next,  we explain the features of the distribution of the sum of $p_z$ for all TI  events, both the experimental and the ECBB model ones (\fig{fig:SOM_TI}), in terms of the direct and  delayed $(e^{\text{-}},2e^\text{-})$ pathways.   
In \fig{fig:SOM_TI_pathways},  for all three intensities, we show that the distribution of  the sum of $p_{z}$    extends up to roughly $\pm 3\times 2\sqrt{U_{p}}$ for the direct pathway (grey arrows in \fig{fig:SOM_TI_pathways}(a)) 
and  up to  $\pm 2\times 2\sqrt{U_{p}}$ for the delayed $(e^{\text{-}},2e^{\text{-}})$ pathway (blue arrows in \fig{fig:SOM_TI_pathways}(a)). This is due to three electrons in the direct and  two electrons in the delayed pathway escaping with large momentum $2\sqrt{U_{p}}$. This is consistent with the distribution of the sum of $p_{z}$
for all TI events extending up to $\pm \beta \times 2\sqrt{U_{p}}$, with $2<\beta<3$. Also, for both pathways, the distributions are doubly peaked giving rise to the double peaks of the  distribution of the sum of $p_{z}$ for all TI events.
Moreover, in the direct pathway the distribution is roughly zero around the sum of  $p_{z}$ being zero. In contrast, in the delayed  pathway, with increasing intensity, the peaks become less pronounced with an increasing probability for the  sum of $p_z$  to be around zero. 
 Hence, this feature observed  in  the distribution of the sum of $p_{z}$ for all TI events (\fig{fig:SOM_TI}) is due to the delayed pathway. For the H-model, we find the direct to be a minor pathway, while the delayed $(e^{\text{-}},2e^{\text{-}})$ one contributes the most to TI.
  This is consistent with soft potentials not accurately describing  scattering of a recolliding electron from the core \cite{Pandit2018}. 

\begin{figure}[t]
\centering
\includegraphics[width=\linewidth]{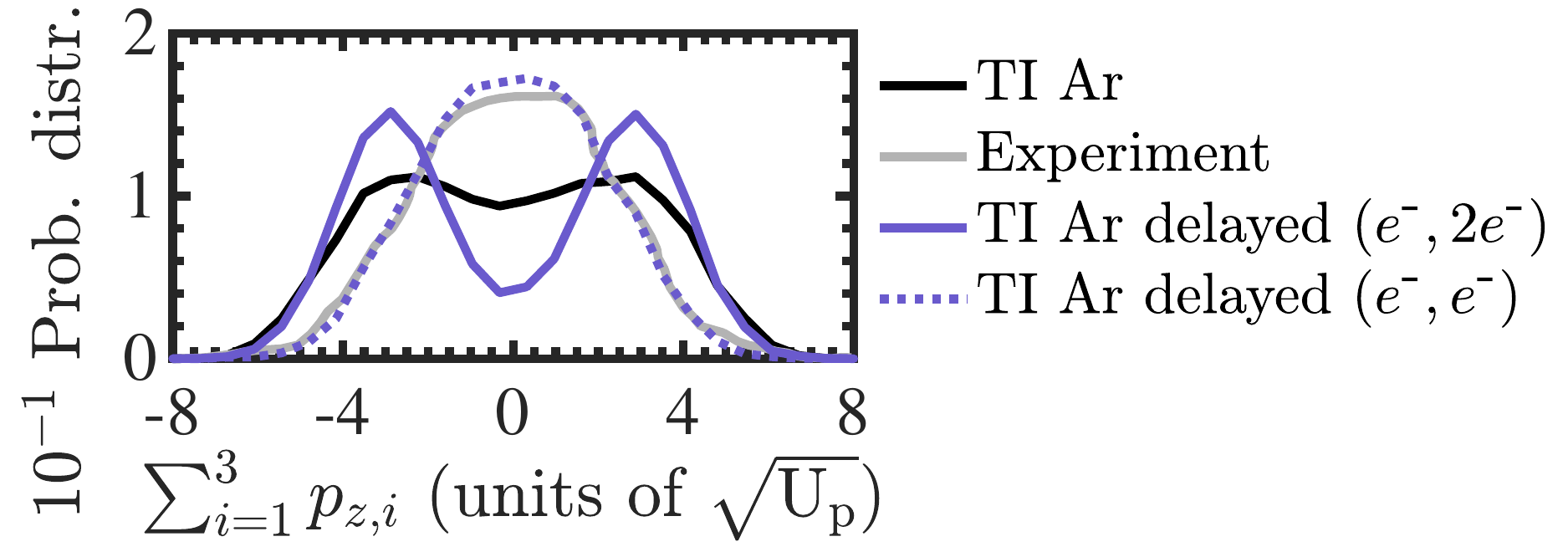}
\caption{For  Ar, 
probability distributions of the sum of   $p_z$ for TI obtained  experimentally \cite{Zrost_2006} (grey) and with the  ECBB model  for all (black) and delayed $(e^{\text{-}},2e^\text{-})$  (blue) and delayed $(e^{\text{-}},e^\text{-})$ events (blue dotted). }\label{fig:Correlated_momenta_TI_pathways_Ar}
\end{figure}

Finally, for Ar, we find that the delayed $(e^{\text{-}},2e^\text{-})$ and  $(e^{\text{-}},e^\text{-})$  pathways  prevail  at  $\mathrm{0.4 \; PW/cm^2 }$. In the latter pathway, the recolliding electron has enough energy to ionize only one electron soon after recollision. 
In the  $(e^{\text{-}},e^\text{-})$ pathway  electrons escape with very small momenta,  see correlated electron momenta in \cite{sup_mat}. The contribution of  these pathways to the distribution of  the sum of $p_{z}$ for all TI events is shown in  \fig{fig:Correlated_momenta_TI_pathways_Ar}. As for TI of Ne, for  Ar, the distribution of the sum of $p_{z}$ corresponding to the  $(e^{\text{-}},2e^\text{-})$ pathway is doubly peaked with a non zero value around the sum of $p_{z}$ being zero (blue line in  \fig{fig:Correlated_momenta_TI_pathways_Ar}). The  distribution  corresponding to the  $(e^{\text{-}},e^\text{-})$ pathway (blue dotted line in  \fig{fig:Correlated_momenta_TI_pathways_Ar})
peaks around  the sum of $p_{z}$ being zero. Interestingly, this distribution of  the delayed $(e^{\text{-}},e^\text{-})$ pathway is in very good agreement with the experimental distribution (grey line)  at $\mathrm{0.3 \; PW/cm^2 }$, with no measurements available at $\mathrm{0.4 \; PW/cm^2 }$. The ECBB model distribution  is more wide compared to the experimental one but also has a significant value around the sum of $p_{z}$ being zero. Hence, for Ar, the ECBB model overestimates the contribution of the more correlated $(e^{\text{-}},2e^\text{-})$ versus the less correlated $(e^{\text{-}},e^\text{-})$ delayed pathway. Given the above, it is clear that  three-electron escape is significantly less correlated in Ar than Ne. 





In conclusion, we demonstrate that  the ECBB 3D semiclassical model is a powerful tool to study correlated multi-electron escape in driven atoms. To do so, we study three-electron ionization in Ne driven by infrared pulses. We show that the triple ionization probability distribution of the sum of the final electron momenta  obtained with the ECBB model is in very good agreement  with experiments. This agreement supports the premise of the ECBB model. That is, to obtain  accurate multi-electron ionization spectra it is important during a recollision to accurately account for the interaction between the recolliding  and the bound electron and for the interactions of  the bound and recolliding electron with the core.   Fully accounting for the singularity in the Coulomb electron-core potential in the ECBB model challenges the generally accepted practice of employing soft core potentials in strong field science.
The ECBB model is developed in a general framework and can thus be easily extended to address correlated escape of more than three electrons in  driven atoms.  It can  also be extended to address  driven molecules. We expect the ECBB model
will be employed to study problems currently out of reach, leading to identifying  novel ultrafast phenomena and to motivating additional experiments in strong field science.


\begin{acknowledgements}
A.E. and G.P.K. acknowledge the EPSRC Grant EP/W005352/1. A. E is grateful to Paul Corkum for fruitful discussions. The authors acknowledge the use of the UCL Myriad High Throughput Computing Facility (Myriad@UCL), and associated support services, in the completion of this work.
\end{acknowledgements}



\bibliographystyle{apsrev4-1}
\bibliography{bibliography}

\end{document}


\title{Supplemental Material}

\subsection{Rate equations}
A simple model for the description of single and sequential double ionization is given by the rate equations
\begin{align}\label{eq:RateEquations}
\begin{split}
\frac{d N_0(t)}{dt} &=-w_{01}(t) N_0(t) \\
\frac{d N_1(t)}{dt} &=w_{01}(t) N_0(t) -w_{12} N_1(t) \\
\frac{d N_2(t)}{dt} &=w_{12}(t) N_1(t) \\
N_{1}\left(t_0\right) &=N_{2}\left(t_0\right) =0, \quad N_0\left(t_0\right)=1,
\end{split}
\end{align}
where $N_0(t),N_1(t)$ and $N_2(t)$ are the time dependent populations of Ne, Ne${}^{+}$ and Ne${}^{2+}$ respectively. The modified    Ammosov-Delone-Krainov (ADK) rate \cite{Landau,Delone:91} developed by Tong and Lin \cite{Tong_2005} describes  the transition from ion state $i$ to ion state $i+1$ with $i=0,1$ as follows
\begin{equation}
w_{i \; i+1}(t)=w_{\mathrm{ADK}}(t) \exp \left[-\beta \frac{Z_c^2 E(t)}{I_p\left(2 I_p\right)^{3 / 2}}\right],
\end{equation}
where $Z_{c}$ is the asymptotic charge  the tunneling electron sees when it tunnels from state $i$, $I_{p}$ is the ionization potential.  The numerical factor $\beta$ is adjusted such that the rate $w_{i \; i+1}(t)$ fits the numerical solution of the Schr\"{o}dinger equation  \cite{Tong_2005}. 
For each transition $i \to i+1$ we use the values for the parameters $Z_c,I_p,\beta$ as listed  in Refs. \cite{Tong_2002,Tong_2005}.


Integrating the rate equation for the ground state, we  obtain the probability  to be in the ground state of Ne as a function of time, i.e the population of this state, 
\begin{equation}\label{eq:N0}
N_0(t) = \exp \left[ - \int_{t_0}^{t} w_{01}(t')dt' \right],
\end{equation}
where we use the initial condition $N_0(t_0)=1$. Solving numerically the other two equations,  we obtain the probability for sequential double ionization, $N_{2}$, at asymptotically large times. That is,  we obtain the probability for each of the two electrons to tunnel ionize due to the lowering of the Coulomb barrier  by the laser field.  For the laser pulse parameters considered in the current work,  at intensity $\mathrm{1.6 \; PW/cm^2 }$ for driven Ne and at $\mathrm{0.4 \; PW/cm^2 }$ for driven Ar, we find that $N_2$ is very small compared to the probability for double ionization we obtain with full scale calculations using the ECBB  model \cite{Agapi3electron}.

Moreover, accounting for the depletion of the initial state, the rate for tunneling from the ground state is obtained as the product of the $w_{01}(t)$ rate  times the probability to be in the ground state, that is, 

\begin{equation}\label{eq:N1}
w_{01}(t)\exp \left[ - \int_{t_0}^{t} w_{01}(t')dt' \right].
\end{equation}
This is the rate we use as the probability distribution in the importance sampling we employ in our full scale calculations with the ECBB model. We account for depletion of the initial ground state for driven Ne and Ar  due to the high intensities considered here and the relatively small ionization potentials $I_{p}$ of Ne and Ar. 

\subsection{Hamiltonian of the four-body system for the H model}

The Hamiltonian of the four-body system for the H model is given by \cite{Agapi3electron}
\begin{equation}\label{Hamiltonian_heis}
H = \sum_{i=1}^{4}\frac{\left[\mathbf{\tilde{p}}_{i}- Q_i\mathbf{A}(\mathbf{r}_{i},t) \right]^2}{2m_i}+\sum_{i=1}^{3}\sum_{j=i+1}^{4}\frac{Q_iQ_j}{|\mathbf{r}_{i}-\mathbf{r}_{j}|} + \sum_{i=2}^{4}V_{H,i}.
\end{equation}
The Heisenberg potential, originally proposed by Kirschbaum and Wilets in Ref. \cite{PhysRevA.21.834}, is given by
\begin{equation}\label{Heisenberg}
V_{H,i}=\dfrac{\xi^2}{4\alpha \mu r_{i,1}^2}\exp \left\{ \alpha \left[ 1 - \left(  \dfrac{r_{i,1} p_{i,1} }{\xi}  \right)^4 \right]  \right\},
\end{equation}
where $ \mathbf{r}_{i,1}=\mathbf{r}_1 - \mathbf{r}_i  $ is the relative position of each one of the three electrons $i=2,3,4$ with respect to the core $i=1$, $\mathbf{p}_{i,1}$ is the corresponding relative momentum
\begin{equation}
\mathbf{p}_{i,1} = \dfrac{m_i\mathbf{p}_1 - m_1\mathbf{p}_i}{m_1 + m_i},
\end{equation}
 and $\mu = m_1m_i/(m_i+m_1) $ is the reduced mass of the electron-core system. 
This potential restricts the relative position and momentum of electron $i$ according to
\begin{equation}\label{Constraint}
r_{i,1}p_{i,1}  \geq \xi.
\end{equation}
The parameter $\alpha$ determines the accessible phase space during the time propagation, with a larger value of  $\alpha$ restricting more the phase space.

\subsection{Definition of quasifree  and bound electron}\label{switching}
In the ECBB model, we fully account for the Coulomb interaction  between a pair of electrons where at least one is quasifree. However, we approximate the interaction between a bound-bound electron pair using    effective Coulomb potentials. The effective potentials chosen ensure that there is no artificial 
autoionization due to our fully accounting for the singularity in the electron-core potential. Hence, to determine whether during time propagation, i.e. on the fly, the interaction between an electron pair is described by the full Coulomb potential or by an effective potential, we need to define on the fly if an electron is quasifree or bound. At the start of propagation, $t_{0}$, we label the tunnel-ionizing electron  (electron 2) as quasifree and the other two (electrons 3 and 4) as bound. We denote the core as particle 1.

\begin{figure}[H]
\centering
\includegraphics[width=\linewidth]{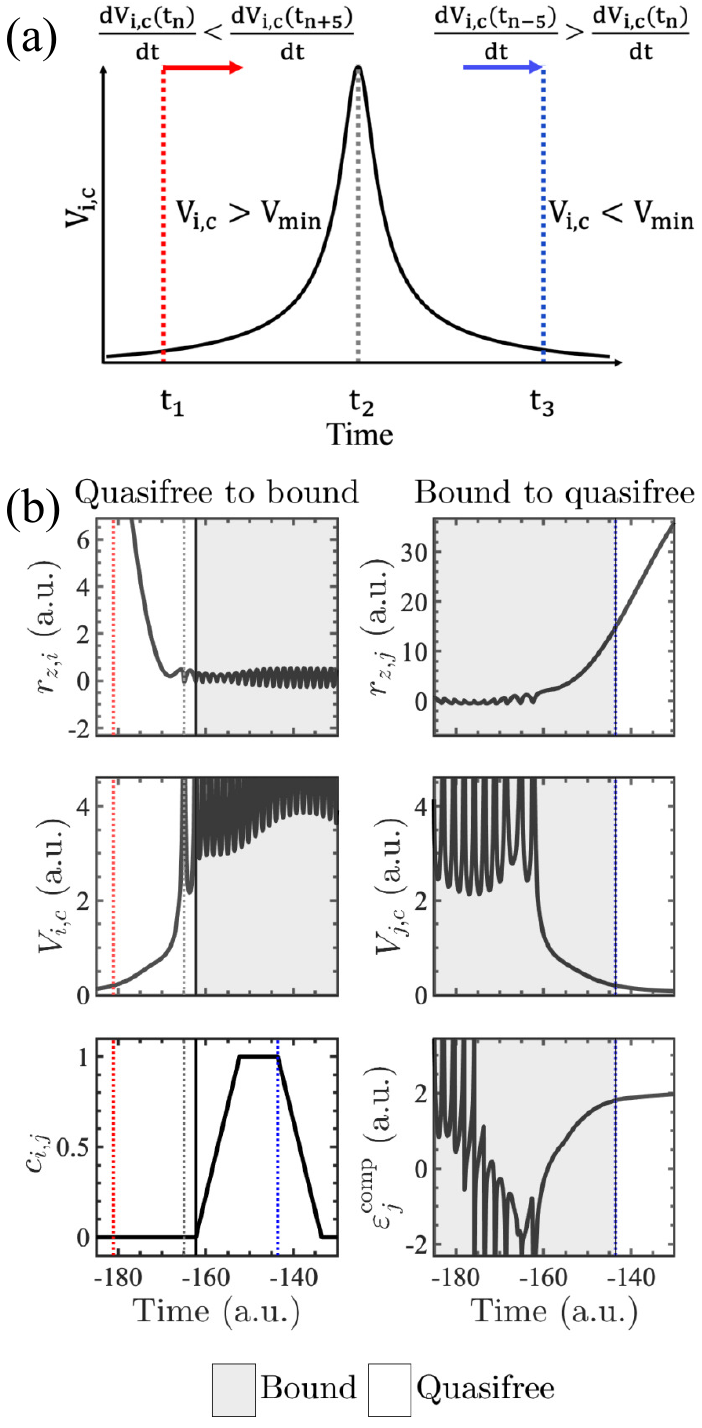}
\caption{Schematic illustration of the criteria to determine when a quasifree electron becomes bound (left column) and when a bound electron becomes quasifree (right column).}
\label{fig:recollision}
\end{figure}

At  times $t>t_0$, a quasifree electron $i$ transitions to bound if the following conditions are satisfied: (i) the potential of electron $i$ with the core, $V_{i,c}$,  is larger than a threshold value, i.e. $V_{i,c}> V_{min}$ at $t_1$, and $V_{i,c}$ is continuously increasing, i.e. $\frac{dV_{i,c}(t_{n+5})}{dt} > \frac{dV_{i,c}(t_{n})}{dt}$ for five times $t_n$ which are five time steps apart with the first one being at time $t_1$, see \fig{fig:recollision}(a); (ii) the position of electron $i$ along the electric field, i.e. $z$ axis here,  has at least two extrema of the same kind, i.e. two maxima or two minima, in a time interval less than half a period of the laser field. Indeed, if the motion of electron $i$ is mostly influenced by the electric field, we expect that its position along the $z$ axis should have only one extremum in a time interval equal to half a period of the laser field.
We start checking if condition (ii) is satisfied at  time $t_2$ when  electron $i$ has the closest approach to the core, i.e. $V_{i,c}$ is maximum. We stop checking whether condition (ii) is satisfied at  time $t_3$ when $V_{i,c}$ is smaller than the threshold value $V_{min}$ and $V_{i,c}$ is continuously decreasing, i.e. $\frac{dV_{i,c}(t_n)}{dt} < \frac{dV_{i,c}(t_{n-5})}{dt}$ for five times $t_n$ which are five time steps apart with the last one being at time $t_3$, see \fig{fig:recollision}(a).   Here, we take $V_{min}$ equal to 3/15 (with 3 the charge of the core) which is equal to 0.2 a.u. We find that our results roughly remain   the same for a range of values of $V_{min}$. 
Also, at the end of the laser pulse, we check whether the compenstated energy  \cite{Leopold_1979} of a  quasifree electron is positive or negative. If the latter occurs, we label the electron as bound.
 Accounting for the effective Coulomb potential of electron $i$ with all other bound electrons, the compensated energy of electron $i$  takes the form
\begin{align}\label{eqn::compensated1}
\varepsilon^{\mathrm{comp}}_{i}(t)= \frac{\mathbf{\tilde{p}}^2_{i} }{2m_i} + \frac{Q_1 Q_i}{|\mathbf{r}_1-\mathbf{r}_i|}  + \sum_{\substack{\;j=2 \\ j \neq i}}^{N} c_{i,j}(t)V_{\mathrm{eff}}(\zeta_j,|\mathbf{r}_{1}-\mathbf{r}_{i}|).
\end{align}

A bound electron transitions to quasifree  at time $t>t_0$ if either one of the following two conditions is satisfied: (i) at time t the compensated energy  of electron $i$ converges to a positive value; (ii) at times $t=t_3$, $V_{i,c}$ is smaller than the threshold value $V_{min}$ and $V_{i,c}$ is continuously decreasing, i.e. $\frac{dV_{i,c}(t_n)}{dt} < \frac{dV_{i,c}(t_{n-5})}{dt}$ for five $t_n$ which are five time steps apart,  the last one being at  $t_3$.

We illustrate the above criteria in \fig{fig:recollision}(b). We  denote the times $t_1$, $t_2$ and $t_3$ with red, grey and blue vertical dashed lines, respectively. In the left column, we plot the position $r_z$ and the potential $V_{i,c}$ of a quasifree electron as it transitions to bound. In the right column, we plot the position $r_z$, the potential $V_{i,c}$ and the compensated energy of a bound electron as it transitions to quasifree. The blue dashed line denotes the time $t_3$ when the electron transitions from bound to quasifree. For this specific trajectory, the time $t_3$ is prior to the time at which the compensated energy would have converged to a positive value.

\subsection{Identifying the pathways of triple  and double ionization events}

We obtain triple ionization (TI) and double ionization (DI) events with a code that incorporates the formulation of the ECBB model, described in  \citep{Agapi3electron}, where we determine on the fly during propagation whether an electron is quasifree or bound. Once we obtain these events, we perform a detailed analysis with a different code, where we  still determine on the fly if an  electron is  quasifree or bound.
 We register a triple or double ionization event as direct if three or two electrons, respectively, ionize shortly after recollision. We register a triple or double ionization event as delayed if one electron ionizes with a delay after recollision. For DI, we have also discussed in previous studies how to identify direct and delayed events \cite{Slingshot}. For TI,
we identify direct  and delayed events as follows:
\begin{enumerate}
\item{We find the ionization time of each electron, $t^i_{\text{ion}},$  for the three electrons in TI and  the two ionizing electrons in DI.}
\item{ We register the maxima in the interelectronic potential energies as a function of time between electron pairs $i,\; j$ and $i,\; k$ and $j,\; k$ during the time intervals when in these pairs one electron is quasifree and the other is bound. Next, for each electron $i$, we identify the maximum for each one of the $i,\; j$ and $i,\; k$ potential energies that is closest to the time $t^i_{\text{ion}}.$ We denote these times as $t_{\text{rec}}^{i,j}$ and $t_{\text{rec}}^{i,k}$. We obtain at most six such times for TI events and four for DI events.}
\item{For each time $t_{\text{rec}}^{i,j}$ we identify the time $t_2$ of closest approach to the core of the quasifree electron (either electron $i$ or $j$)  that is closest to $t_{\text{rec}}^{i,j}$ and denote it as $t^{i,j}_2$. We obtain at most six such times for TI events and four for DI events.}
\end{enumerate}
We label an event as direct or delayed TI if four of the times $t^{i,j}_2$ are the same, accounting for one electron being quasifree and the other two bound. That is, if electron $i$ is quasifree during the recollision closest to the ionisation time $t^i_{\text{ion}}$ then the times $t^{i,j}_2$, $t^{i,k}_2$, $t^{j,i}_2$ and $t^{k,i}_2$ should be the same. The times $t^{j,i}_2$ and $t^{k,i}_2$ are associated with the recollision times $t^{j,i}_{\text{rec}}$ and $t^{k,i}_{\text{rec}}$ for the bound electrons $j$ and $k$, respectively. For the quasifree electron we obtain two recollision times $t^{i,j}_{\text{rec}}$ and $t^{i,k}_{\text{rec}}$ associated with the ionization time $t^i_{\text{ion}}$. We choose  the one with the largest difference from  $t^i_{\text{ion}}$, guaranteeing a stricter criterion for direct TI events. 
 Next, we label a TI event as direct  if the following  conditions are satisfied $\Delta t_{1}=|t^{i,j}_{\text{rec}} - t^i_{\text{ion}}| < t_{\text{diff}}$ or $(t^i_{\text{ion}} < t^{i,j}_{\text{rec}} \; \text{and} \; t^i_{\text{ion}} < t^{i,k}_{\text{rec}})$ and $\Delta t_{2}=|t^{j,i}_{\text{rec}} - t^j_{\text{ion}}| < t_{\text{diff}}$ and $\Delta t_{3}=|t^{k,i}_{\text{rec}} - t^k_{\text{ion}}| < t_{\text{diff}}$. The condition $(t^i_{\text{ion}} < t^{i,j}_{\text{rec}} \; \text{and} \; t^i_{\text{ion}} < t^{i,k}_{\text{rec}})$ has also been used in our previous studies \cite{ChenA2017Ndiw,Slingshot} to account for a quasifree electron ionising significantly earlier before recollision. This happens mostly at high intensities.   We label events as delayed pathway TI, with only one electron ionizing shortly after recollision,  if two out of the three times $\Delta t_{1}$, $\Delta t_{2}$ and $\Delta t_{3}$ are larger than $t_{\text{diff}}$ and one is less than $t_{\text{diff}}$.
  A similar process is followed to identify  direct and delayed  DI events.
The time  $t_{\text{diff}}$ is determined by the time interval where the interelectronic potential energy undergoes a sharp change due to a recollision. For the intensities considered here, we find  $t_{\text{diff}} \approx T/8$, with $T$ being the period of the laser field.


\subsection{Double ionization probabilities for the H model}
\begin{table}[H]
\centering
\begin{tabular}{lclclcl}
\hline
\hline
\multicolumn{3}{c} {Double ionization probability (\%)} \\
\hline
Intensity ($\mathrm{PW/cm^2}$)& $\alpha = 2$ & $\alpha = 4$\\
\hline
1.0 & 0.7 & 0.4 \\  
1.3 & 0.8 & 0.6 \\ 
1.6 & 1.5 & 2.3 \\ 
\hline
\hline
\end{tabular}
\caption{Probabilities for double ionization  obtained using the H model for Ne interacting with a laser pulse at 800 nm and 25 fs duration for $\alpha=2$ and $\alpha=4$.}\label{Tab:DI_prob}
\end{table}

In Table \ref{Tab:DI_prob}, we show the probability for double ionization of Ne driven by a laser pulse at 800 nm and  25 fs duration, at intensities 1.0, 1.3 and 1.6 $\mathrm{PW/cm^2}$, obtained  using the H model with $\alpha =2$ and $\alpha = 4.$ We find that the  probability for double ionization depends on  $\alpha.$

\subsection{Correlated momenta}

\begin{figure}[H]
\centering
\includegraphics[width=\linewidth]{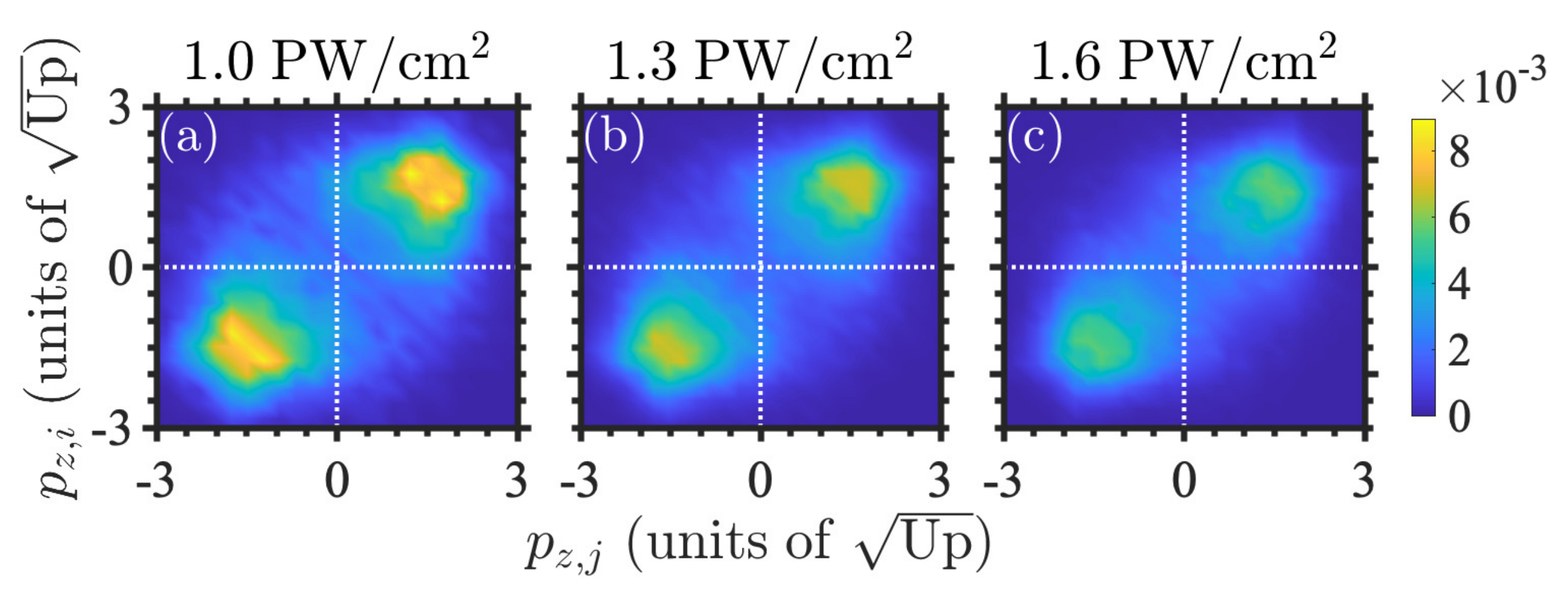}
\caption{Correlated momenta of all three pairs of escaping electrons for triple ionization of driven Ne for the three intensities under consideration. }\label{fig:Cor_mom_TI_Neon}
\end{figure}

In \fig{fig:Cor_mom_TI_Neon}, we show the correlated momenta for all three electron pairs in triple ionization of driven Ne for intensities 1.0, 1.3 and 1.6 $\mathrm{PW/cm^2}$. We find that the three electrons escape with large momenta. This is consistent with the direct pathway being one of the two 
prevailing pathways of triple ionization for driven Ne.

\begin{figure}[H]
\centering
\includegraphics[width=\linewidth]{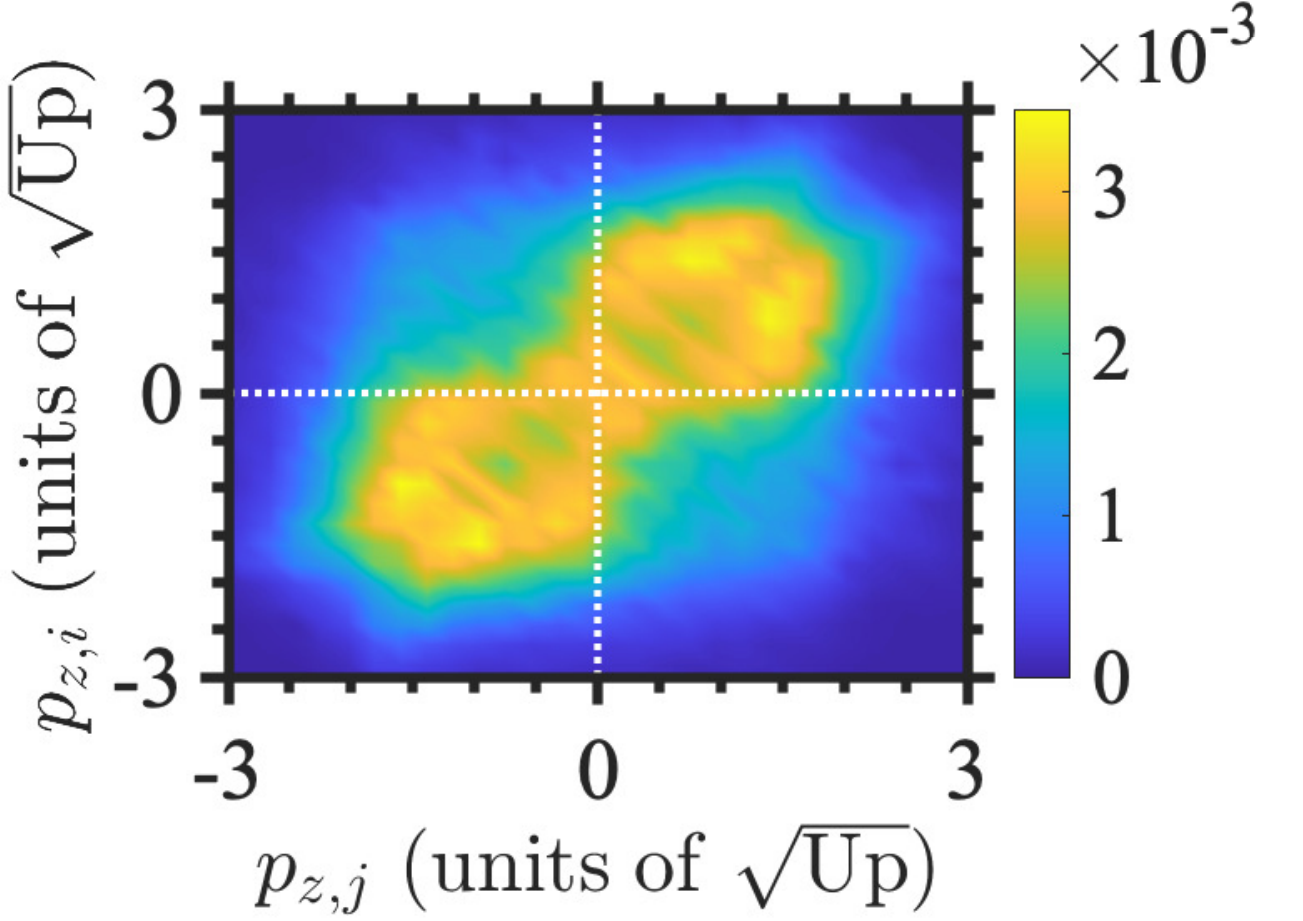}
\caption{Correlated momenta of all three pairs of escaping electrons for triple ionization of driven Ar.}\label{fig:Cor_mom_TI_Argon}
\end{figure}

In \fig{fig:Cor_mom_TI_Argon}, we show the correlated momenta for all three electron pairs in triple ionization of driven Ar for intensity 0.4 $\mathrm{PW/cm^2}$. We find that the three electrons escape with smaller momenta than for driven Ne. This is consistent with the delayed pathways 
prevailing triple ionization of driven Ar.

\FloatBarrier

\bibliographystyle{apsrev4-1}
\bibliography{bibliography}{}